\documentclass[11pt,superscriptaddress]{revtex4-1}
\usepackage{amsmath,amssymb}
\usepackage{graphicx}
\usepackage{float}
\usepackage{subfigure}
\usepackage{verbatim}
\usepackage{lineno}
\usepackage{amsfonts}
\usepackage{dcolumn,tabularx}
\usepackage{bm}
\usepackage{color}
\usepackage[colorlinks,citecolor=blue]{hyperref}

\begin{document}
\hyphenpenalty=5000
\tolerance=1000

\def\linenumberfont{\normalfont\small}
\switchlinenumbers
\pagestyle{plain}
\pagenumbering{arabic}

\title{Room-temperature coherent control of implanted defect spins in silicon carbide}

\affiliation{Synergetic Innovation Center of Quantum Information and Quantum Physics, University of Science and Technology of China, Hefei 230026, People's Republic of China}

\affiliation{State Key Laboratory of Functional Materials for Informatics, Shanghai Institute of Microsystem and Information Technology, Chinese Academy of Sciences, Shanghai, 200050, People's Republic of China}

\affiliation{CAS Key Laboratory of Quantum Information, University of Science and Technology of China, Hefei 230026, People's Republic of China}

\affiliation{Center of Materials Science and Optoelectronics Engineering, University of Chinese Academy of Sciences, Beijing, 100049, People's Republic of China}

\affiliation{CAS Key Laboratory of Microscale Magnetic Resonance and Department of Modern Physics, University of Science and Technology of China, Hefei 230026, China.}

\affiliation{Hefei National Laboratory for Physical Sciences at the Microscale, University of Science and Technology of China, Hefei 230026, China.}

\affiliation{Department of Atomic Physics, Budapest University of Technology and Economics, Budafoki \'ut. 8, H-1111, Hungary}

\affiliation{Wigner Research Center for Physics, PO. Box 49, H-1525, Hungary}

\author{Fei-Fei Yan}
\thanks{These authors contributed equally to this work.}
\affiliation{Synergetic Innovation Center of Quantum Information and Quantum Physics, University of Science and Technology of China, Hefei 230026, People's Republic of China}
\affiliation{CAS Key Laboratory of Quantum Information, University of Science and Technology of China, Hefei 230026, People's Republic of China}

\author{Ai-Lun Yi}
\thanks{These authors contributed equally to this work.}
\affiliation{State Key Laboratory of Functional Materials for Informatics, Shanghai Institute of Microsystem and Information Technology, Chinese Academy of Sciences, Shanghai, 200050, People's Republic of China}
\affiliation{Center of Materials Science and Optoelectronics Engineering, University of Chinese Academy of Sciences, Beijing, 100049, People's Republic of China}

\author{Jun-Feng Wang}
\affiliation{Synergetic Innovation Center of Quantum Information and Quantum Physics, University of Science and Technology of China, Hefei 230026, People's Republic of China}
\affiliation{CAS Key Laboratory of Quantum Information, University of Science and Technology of China, Hefei 230026, People's Republic of China}

\author{Qiang Li}
\affiliation{Synergetic Innovation Center of Quantum Information and Quantum Physics, University of Science and Technology of China, Hefei 230026, People's Republic of China}
\affiliation{CAS Key Laboratory of Quantum Information, University of Science and Technology of China, Hefei 230026, People's Republic of China}

\author{Pei Yu}
\affiliation{Synergetic Innovation Center of Quantum Information and Quantum Physics, University of Science and Technology of China, Hefei 230026, People's Republic of China}
\affiliation{CAS Key Laboratory of Microscale Magnetic Resonance and Department of Modern Physics, University of Science and Technology of China, Hefei 230026, China.}
\affiliation{Hefei National Laboratory for Physical Sciences at the Microscale, University of Science and Technology of China, Hefei 230026, China.}

\author{Jia-Xiang Zhang}
\affiliation{State Key Laboratory of Functional Materials for Informatics, Shanghai Institute of Microsystem and Information Technology, Chinese Academy of Sciences, Shanghai, 200050, People's Republic of China}
\affiliation{Center of Materials Science and Optoelectronics Engineering, University of Chinese Academy of Sciences, Beijing, 100049, People's Republic of China}

\author{Adam Gali}
\affiliation{Department of Atomic Physics, Budapest University of Technology and Economics, Budafoki \'ut. 8, H-1111, Hungary}
\affiliation{Wigner Research Center for Physics, PO. Box 49, H-1525, Hungary}

\author{Ya Wang}
\affiliation{Synergetic Innovation Center of Quantum Information and Quantum Physics, University of Science and Technology of China, Hefei 230026, People's Republic of China}
\affiliation{CAS Key Laboratory of Microscale Magnetic Resonance and Department of Modern Physics, University of Science and Technology of China, Hefei 230026, China.}
\affiliation{Hefei National Laboratory for Physical Sciences at the Microscale, University of Science and Technology of China, Hefei 230026, China.}

\author{Jin-Shi Xu}\email{jsxu@ustc.edu.cn}
\affiliation{Synergetic Innovation Center of Quantum Information and Quantum Physics, University of Science and Technology of China, Hefei 230026, People's Republic of China}
\affiliation{CAS Key Laboratory of Quantum Information, University of Science and Technology of China, Hefei 230026, People's Republic of China}

\author{Xin Ou}\email{ouxin@mail.sim.ac.cn}
\affiliation{State Key Laboratory of Functional Materials for Informatics, Shanghai Institute of Microsystem and Information Technology, Chinese Academy of Sciences, Shanghai, 200050, People's Republic of China}
\affiliation{Center of Materials Science and Optoelectronics Engineering, University of Chinese Academy of Sciences, Beijing, 100049, People's Republic of China}

\author{Chuan-Feng Li}\email{cfli@ustc.edu.cn}
\affiliation{Synergetic Innovation Center of Quantum Information and Quantum Physics, University of Science and Technology of China, Hefei 230026, People's Republic of China}
\affiliation{CAS Key Laboratory of Quantum Information, University of Science and Technology of China, Hefei 230026, People's Republic of China}

\author{Guang-Can Guo}
\affiliation{Synergetic Innovation Center of Quantum Information and Quantum Physics, University of Science and Technology of China, Hefei 230026, People's Republic of China}
\affiliation{CAS Key Laboratory of Quantum Information, University of Science and Technology of China, Hefei 230026, People's Republic of China}

\date{\today }

\begin{abstract}
Recently, vacancy-related spin defects in silicon carbide (SiC) have been demonstrated to be potentially suitable for versatile quantum interface building and scalable quantum network construction. Significant efforts have been undertaken to identify spin systems in SiC and to extend their quantum capabilities using large-scale growth and advanced nanofabrication methods. Here we demonstrated a type of spin defect in the 4H polytype of SiC generated via hydrogen ion implantation with high-temperature post-annealing, which is different from any known defects. These spin defects can be optically addressed and coherently controlled even at room temperature, and their fluorescence spectrum and optically detected magnetic resonance spectra are different from those of any previously discovered defects. Moreover, the generation of these defects can be well controlled by optimizing the annealing temperature after implantation. These defects demonstrate high thermal stability with coherently controlled electron spins, facilitating their application in quantum sensing and masers under harsh conditions.\\

{\bf Keywords}\ \ \ ODMR, SiC, spin defects, Room-temperature, coherent control
\end{abstract}

\maketitle

\section{Introduction}

Previously, deep defects in semiconductors were considered harmful elements that needed to be eliminated. Nowadays, with the development of quantum information, several optically addressable defect spins have played important roles in optical/spin interface. Some of these defect spins can be individually and coherently controlled at room temperature. Among these, the negatively charged nitrogen-vacancy (NV) center in diamond is a leading platform through which significant achievements have been accomplished in quantum communications \cite{hensen2015loophole}, quantum computation \cite{zu2014experimental,bradley201910,shi2010room}, and quantum sensing \cite{toyli2012measurement,dolde2011electric,kucsko2013nanometre,taylor2008high}. However, scalable quantum network construction for NV centers in diamond still encounter considerable challenges \cite{atature2018material}. To further extend such defect-based applications, many attempts have been made to explore similar defects in other materials; vacancy-related defects in silicon carbide (SiC) are attractive candidates \cite{falk2013polytype,christle2017isolated,klimov2015quantum,gali2011time,
christle2015isolated,zhou2017self,seo2016quantum,whiteley2019spin,koehl2011room,
yan2018coherent}.

As SiC is a promising wide-bandgap semiconductor material with well-established inch-scale growth and device-engineering protocols, it is widely used in high-power electronic devices. High-quality bulk and epitaxial single crystals of SiC can be commercially purchased. There are many different polytypes for SiC with 3C-SiC, 4H-SiC, and 6H-SiC symmetries; 4H-SiC is the most common among them \cite{falk2013polytype}. Similar to NV centers in diamond, certain defects in SiC, such as divacancies (PL1 to PL7) and silicon vacancies (V$_\text{Si}$) in 4H-SiC, can be optically identified and coherently controlled at room temperature \cite{koehl2011room,yan2018coherent,widmann2015coherent}. Furthermore, their near-infrared (NIR) fluorescence signals can be used for biological sensing \cite{somogyi2012near} and optical fiber transmission. Moreover, the Hahn echo coherence time of divacancy defect spins in 4H-SiC has been demonstrated to be 1.3 ms, i.e., the longest coherence time in a naturally isotopic crystal \cite{seo2016quantum}. Apart from intrinsic defects, different implanted methods have been used to generate different types of defects on-demand in SiC with similar optical and spin properties \cite{koehl2017resonant,wang2019demand,von2016nv,spindlberger2019optical,bosma2018identification}. Defect spins in SiC have important applications in many quantum information tasks \cite{christle2017isolated,klimov2015quantum,weber2010quantum}.

Recently, significant efforts have been undertaken to investigate more spin systems embedded in SiC to extend their quantum functionalities \cite{falk2013polytype,christle2017isolated,weber2010quantum}. Herein, we demonstrated new optically addressable paramagnetic point spin defects in the 4H-SiC epitaxy layer generated by implanting 170-keV hydrogen ions (H$^+$) with a fluence of up to 5 $\times 10^{16}$ cm$^{-2}$ and annealing at 1,300$^\circ$C for 2 h (see Methods for more details). The NIR zero-phonon line (ZPL) of defects is at 1,007 nm, which is suitable for biological sensing and located between the first and second communication windows of the fiber. According to the optically detected magnetic resonance (ODMR) spectrum, the ground-state structure of the defect spins is classified as a $c$-axis three-level system. The ODMR contrast of these ensemble defects is similar to that of the divacancy in SiC (reported to be $\sim$ 0.7$\%$), but the resonant frequencies are different from those of any known defects (see the comparison in Supplementary Table \uppercase\expandafter{\romannumeral1} in Supplementary Information (SI) \cite{SI}). The density of the generated defects is largely dependent on the fluence of the implanted ions and annealing temperature. The defects are further shown to be controllably generated by optimizing the annealing temperature. By increasing the implanted fluence, the concentration of these defects (similar to divacancies, the defects are  labeled PL8 according to the photoluminescence (PL) spectrum) can be further increased to a sufficiently high level to be useful for high-sensitivity quantum sensing \cite{degen2017quantum} and maser development \cite{breeze2018continuous}.

\section{Results}

\begin{figure}[!h]
  \centering
  \includegraphics[width=0.6\textwidth]{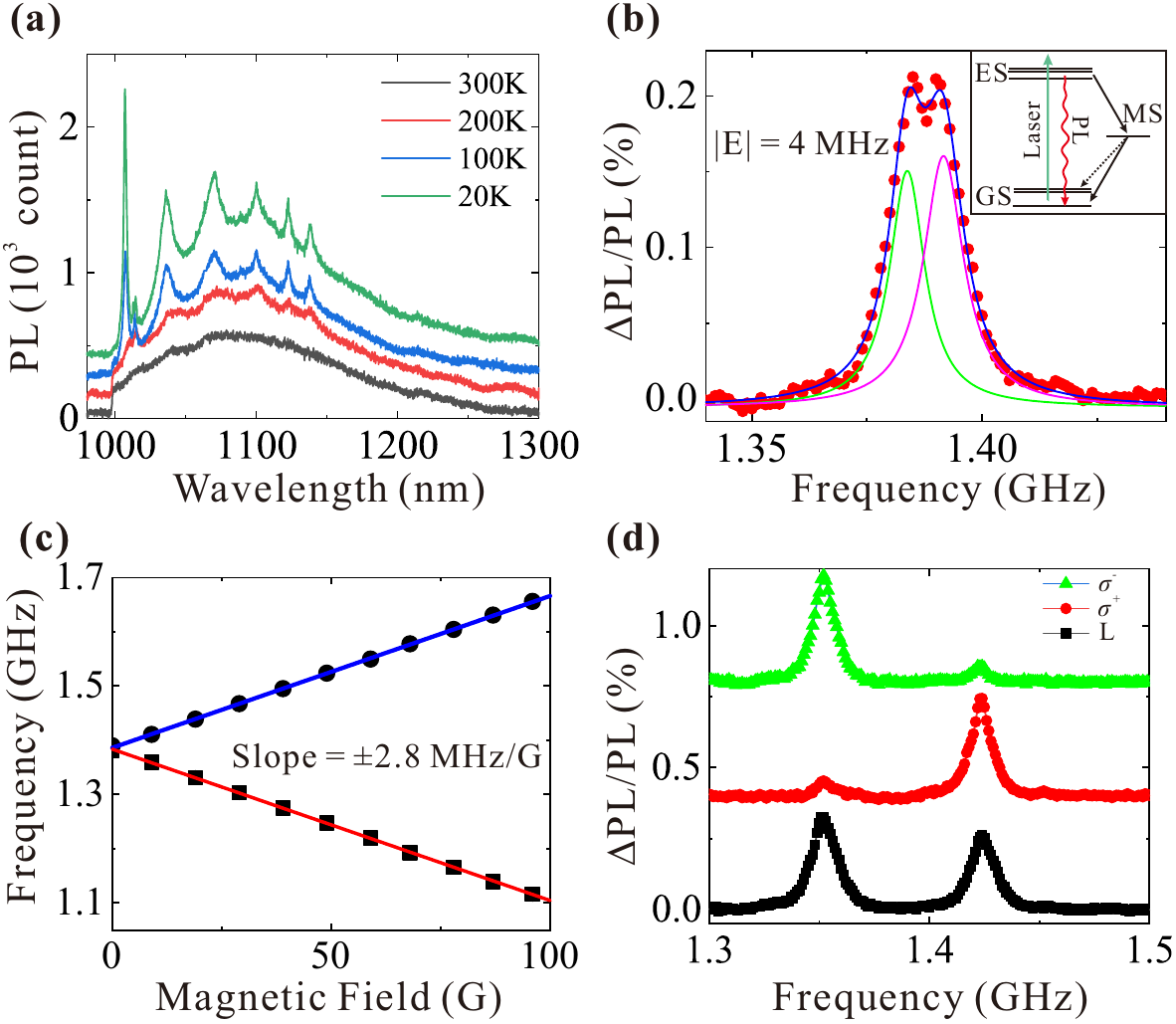}\\

\caption{\textbf{Optical and ODMR spectra of implanted defects (PL8) in 4H-SiC.}  (a) Photoluminescence spectra of PL8 defects at temperatures 20 K$-$300 K with 1,000-nm longpass filter. The zero-phonon line (ZPL) at 1,007 nm belongs to PL8 defects. (b) ODMR spectrum of PL8 defects at room temperature with zero magnetic field. The signal was fitted by two Lorentz functions, each of which is centered at 1,383.8 and 1,391.8 MHz. The zero-field parameters $|D|$ and $|E|$ were 1,387.8 and 4 MHz, respectively. The possible energy level is shown in the inset (GS, ES, and MS refer to the ground, excited, and metastable states, respectively). (c) Zeeman splitting at different magnetic fields. The slope was $\pm 2.8$ MHz/G, implying that the orientation of PL8 defect spins is the $c$-axis. Error bars represent the standard deviations of the corresponding fitting values, which are smaller than the symbols. (d) ODMR spectra pumped with different polarized microwaves at 12.8-G external magnetic field. The green triangles, red dots, and black squares represent the experimental results when the microwave was clockwise circular polarized ($\sigma^{-}$), anticlockwise polarized ($\sigma^{+}$), and linearly polarized (L), respectively.}

\label{novel_center}

\end{figure}

The photoluminescence spectra of the sample at temperatures 20 K$-$300 K are shown in Fig. \ref{novel_center}(a). The sample was excited using a 920-nm laser (see Methods for details), and the peak at 1,007 nm of the low temperature spectra belonged to ZPL of PL8 defects. Several other peaks in the spectrum were deduced to be unrelated to PL8 defects (see Supplementary Fig. 1 in the SI for more detail \cite{SI}). Fig. \ref{novel_center}(b) displays the ODMR spectrum without a magnetic field applied at room temperature that was fitted using two Lorentz functions. Two obvious ODMR peaks centered at 1,383.8 and 1,391.8 MHz are shown. They were different from the ODMR signals of the divacancy defects in SiC \cite{falk2013polytype,koehl2011room,SI}. Based on the ODMR spectrum's features, the ground-state electronic structure was inferred to be a three-level system (shown in the inset of Fig. \ref{novel_center}(b)). The spin ground-state Hamiltonian under the external magnetic field ($\vec{\textbf{B}}_0$) can be described as $H = D \left[ S^2_z - \frac{S(S+1)}{3}\right] +E(S^2_x - S^2_y) + g\mu _B \vec{\textbf{B}}_0 \cdot \vec{\textbf{S}}$, where $D$ and $E$ are the zero-field-splitting parameters and $S_x$, $S_y$ and $S_z$ are the three components of the spin ($\vec{\textbf{S}}$). From the Lorenz fittings in Fig. \ref{novel_center}(b), the zero-field parameters $|D|$ = 1387.8 MHz and $|E|$ = 4 MHz were obtained at room temperature. To confirm that the ground state of PL8 defects were the $c$-axis symmetry, the ODMR signals were measured as a function of the $c$-axis magnetic field. As shown in Fig. \ref{novel_center}(c), the two transitions of PL8 defects were measured to split at a slope of $\pm 2.8$ MHz/G, demonstrating the $c$-axis spin orientation. The property was similar to that of PL6 defects in 4H-SiC \cite{falk2013polytype,yan2018coherent}. Further, we detected the ODMR signals with polarized microwaves implemented using two crossed copper wires with a 90$^\circ$ phase difference \cite{london2014strong} at 12.8-G external magnetic field. The experimental results are shown in Fig. \ref{novel_center}(d). The higher resonant frequency represented the transition between $m_s = 0$ and $m_s = +1$, while the lower resonant frequency represented the transition between $m_s = 0$ and $m_s = -1$. Comparing the transition intensities with a clockwise circular polarized microwave ($\sigma^{-}$, green triangles, transition mainly occurred for $\Delta m_s = -1$), anticlockwise circular polarized microwave ($\sigma^{+}$, red dots, transition mainly occurred for $\Delta m_s = +1$), and linearly polarized microwave (L, black squares, both transitions occurred almost equally), we concluded that PL8 defect spins were primarily initialized to $m_s = 0$, with 920-nm laser pumping (similar to PL6 defects).

\begin{figure}[!h]
  \centering
  \includegraphics[width=0.6\textwidth]{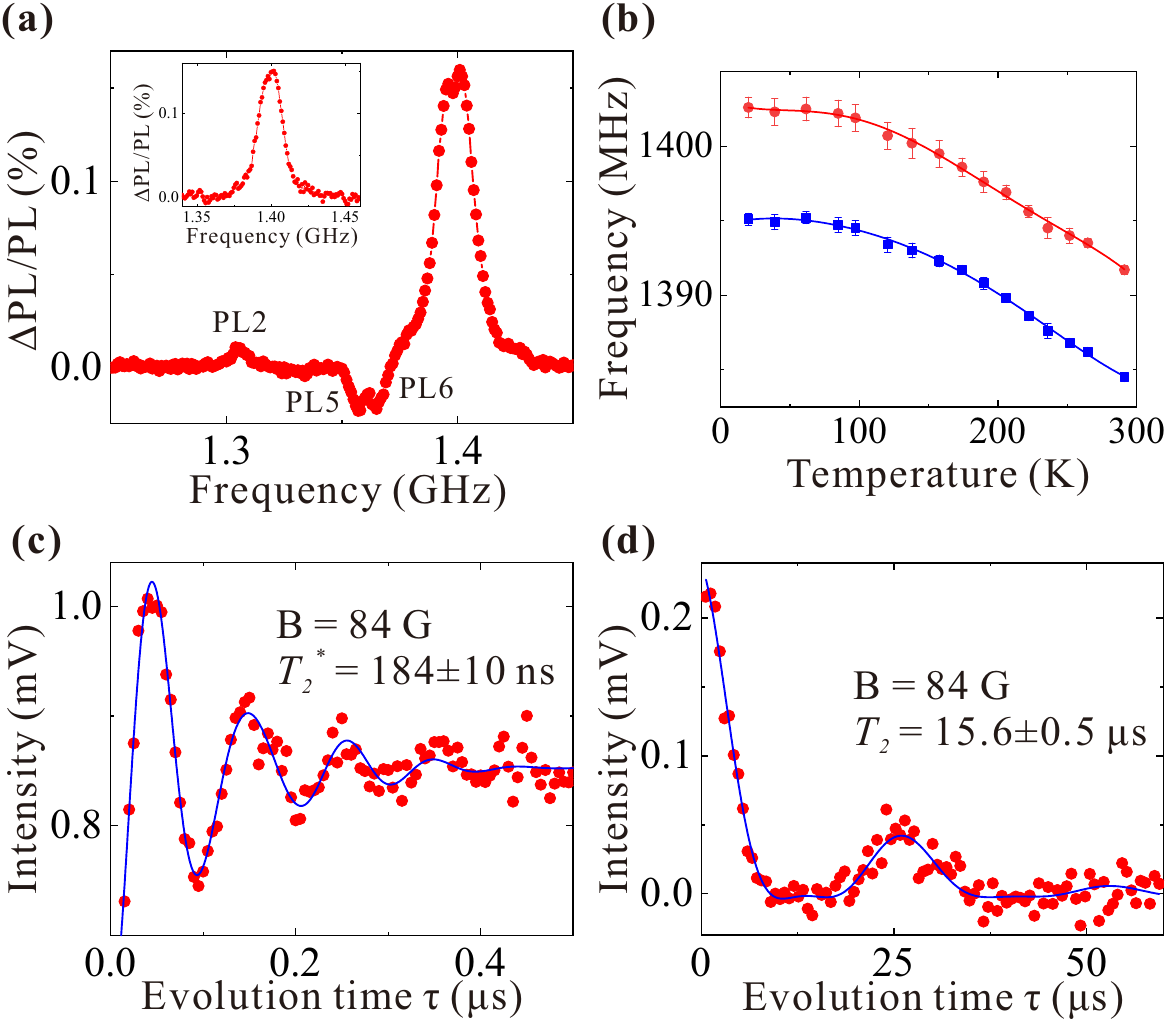}\\

\caption{\textbf{Coherent manipulation of PL8 defect spins at 20 K.} (a) ODMR signal at 20 K without a magnetic field. Signals with significantly lower intensities from other defects of PL2 (1,305.4 MHz), PL5 (1,356.9 MHz), and PL6 (1,365.4 MHz) were also detected. The inset shows ODMR spectrum when only ZPL of PL8 defects was collected. (b) ODMR frequency at zero magnetic field of PL8 defects as a function of temperature. The red dots and blue squares represent the centers of the two corresponding Lorentz fittings, with the red and blue lines representing the fifth-order polynomial fittings; $|D|$ decreased as the temperature increased, while $|E|$ remained unchanged. Error bars represent the standard deviations of the corresponding fitting values. (c) The Ramsey fringe at 84 G, revealing that $T_{2}^{*} = 184 \pm 10$ ns. The microwave detuning was set as $\delta f = 10$ MHz. (d) The optically detected Hahn echo at 84 G, with $T_{2} = 15.6 \pm 0.5$ $\mu$s. The blue lines represent the fittings.}

\label{coherent control at 20K}

\end{figure}

Moreover, we investigated the coherent manipulation of PL8 defect spins at 20 K. Fig. \ref{coherent control at 20K}(a) shows the ODMR spectrum obtained by collecting all photoluminescence from the sample using a 1,000-nm longpass filter. Several weak ODMR signals of other previously reported defects (PL2, PL5, and PL6) were also observed \citep{falk2013polytype}. The ODMR spectrum was further detected by only collecting the photoluminescence around ZPL at 1,007 nm with a band-pass filter from 1,000 to 1,024 nm. The result is shown in the inset of Fig. \ref{coherent control at 20K}(a). Only the signal of PL8 defect spins was observed, confirming ZPL at 1,007 nm. The resonant frequency of the ODMR spectrum at 20 K increased by 10 MHz compared with that at room temperature, as clearly shown in Fig. \ref{coherent control at 20K}(b). This behavior was similar to that of the divacancy in SiC \citep{zhou2017self} and NV centers in diamond \citep{toyli2012measurement}. The red dots and blue squares in the figure represent the centers of the corresponding Lorentz fittings at different temperatures, and the red and blue lines represent fifth-order polynomial fittings \citep{zhou2017self}. Similar to that of PL5 defects in SiC \citep{zhou2017self}, when the temperature increased, the zero-field parameter $|E|$ remained the same. A Ramsey pulse sequence \citep{koehl2011room} was used to measure the free induction decay time $T_2^*$. Fig. \ref{coherent control at 20K}(c) shows the Ramsey oscillation at 84 G with a detuning frequency $\delta f = 10$ MHz, which revealed an inhomogeneous spin coherence time $T_2^* = 184 \pm 10$ $\text{ns}$ for ground-state spins. The fit function was a power-dependent attenuation oscillation function \citep{toyli2012measurement}. A Hahn echo pulse was implemented to detect the homogenous spin coherence time \citep{koehl2011room}; the result is shown in Fig. \ref{coherent control at 20K}(d). From the oscillation decay fitting \citep{koehl2011room}, we obtained $T_2 = 15.6 \pm 0.5$ $\mu \text{s}$. The oscillation was driven by nuclear spins ($^{13}$C and $^{29}$Si).

\begin{figure}[!h]
  \centering
  \includegraphics[width=0.6\textwidth]{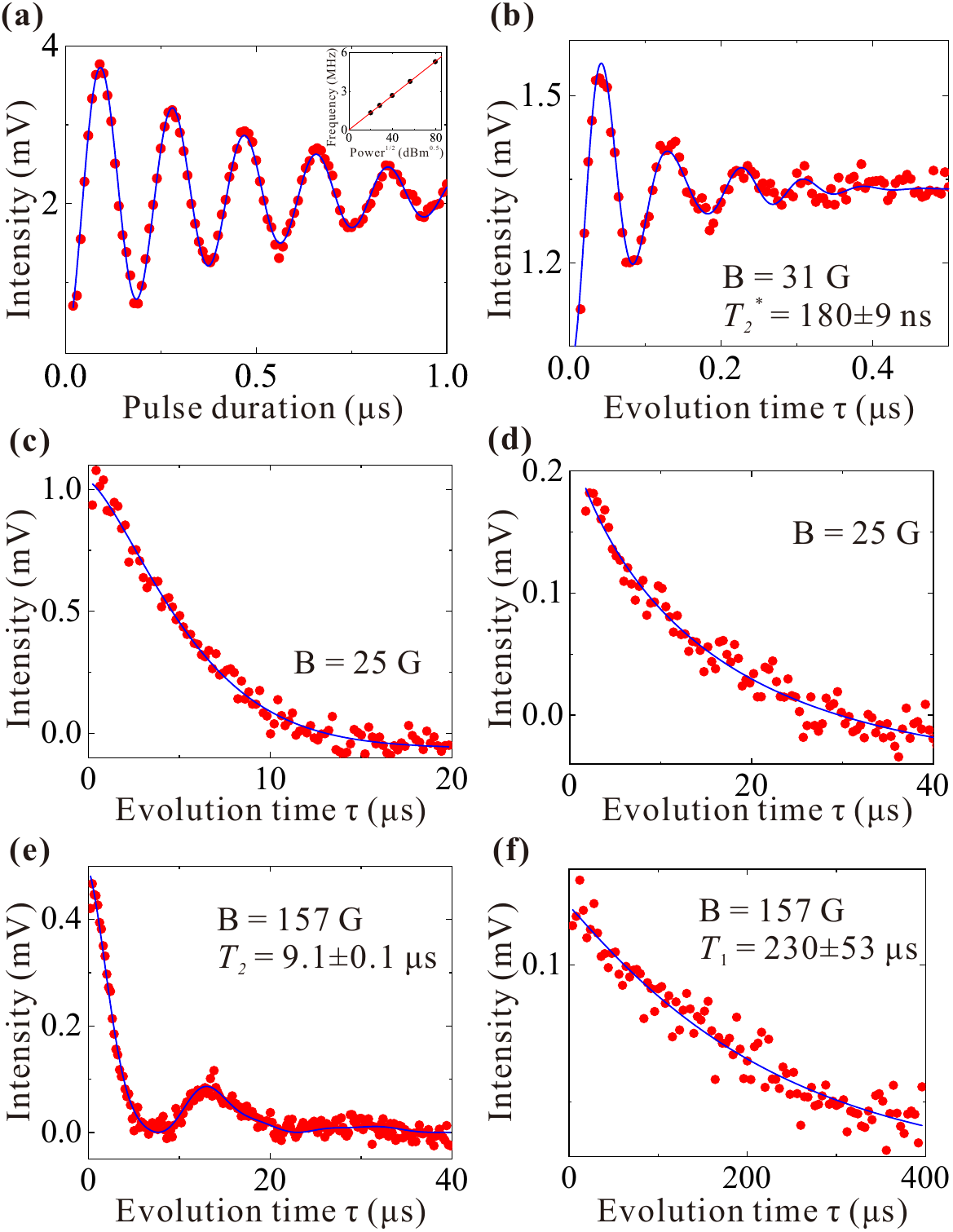}\\

\caption{\textbf{Coherent manipulation of PL8 defect spins at room temperature.} (a) Rabi oscillation observed at 31 G. The inset shows that the Rabi frequency increased with the square root of the microwave power. The red line represents a linear fitting. (b) Ramsey oscillation at 31 G, with $T_{2}^{*}=180\pm 9$ ns. (c) Hahn echo decay at 25 G, revealing the decay time of $6.1 \pm 0.1$ $\mu$s. (d) Carr-Purcell-Meiboom-Gill decay at 25 G with the decay time of $14.7 \pm 1.5$ $\mu$s. (e) The Hahn echo decay at 157 G, with $T_2 = 9.1 \pm 0.1$ $\mu$s. (f) $T_1$  measurement, revealing the longitudinal spin relaxation time $T_1 = 230 \pm 53$ $\mu $s. The $y$-axis represents the voltage intensity in the lock-in amplifier. The blue lines represent the theoretical fittings.}

\label{coherent control at room temperature}

\end{figure}

PL8 defect spins can also be coherently controlled at room temperature. Fig. \ref{coherent control at room temperature}(a) shows the Rabi oscillation at room temperature. The Rabi frequency was demonstrated to be proportional to the square root of the microwave power in the copper wire \citep{widmann2015coherent} (inset of Fig. \ref{coherent control at room temperature}(a)). The Ramsey oscillation was also measured at room temperature; the result is shown in Fig. \ref{coherent control at room temperature}(b). The inhomogeneous spin coherence time $T_2^* = 180 \pm 9$ ns was deduced from the fitting. Compared with the result shown in Fig. \ref{coherent control at 20K}(c), $T_2^*$ was nearly constant and independent of the sample temperature. Fig. \ref{coherent control at room temperature}(c) shows the room temperature spin echo decay time of $6.1 \pm 0.1$ $\mu$s at 25-G external magnetic field. The result was fitted with two frequency components equal to Larmor frequencies of $^{13}$C and $^{29}$Si. The decoherence effect can be suppressed using the dynamical decoupling method. In Fig. \ref{coherent control at room temperature}(d), the Carr-Purcell-Meiboom-Gill (CPMG)-3$\pi$ pulse is implemented \citep{koehl2011room,de2010universal}. Compared with 6.1-$\mu$s decay time in Fig. \ref{coherent control at room temperature}(c), the decay time here was extended to 14.7 $\mu$s at the same magnetic field. The decay time in Fig. \ref{coherent control at room temperature}(c) was shorter than $T_2$ owing to the perturbation of the nuclear spin environment at the magnetic field. We further increased the magnetic field \citep{seo2016quantum}, and the homogenous spin coherence time was measured (Fig. \ref{coherent control at room temperature}(c)) at 157 G, revealing that $T_2 = 9.1 \pm 0.9$ $\mu$s at room temperature, which is longer than that of the implanted PL1 defects in SiC ($T_2 = 6$ $ \mu$s) \citep{falk2013polytype}. Compared with Fig. \ref{coherent control at 20K}(d), $T_2$ at a lower temperature was longer than that at room temperature. We also measured $T_2$ of the implanted PL6 defects for $T_2 = 9.7 \pm 0.8$ $\mu$s (see Supplementary Fig. 3 in SI \cite{SI}), which was almost the same as that of PL8 defects. However, $T_2$ of PL6 in the sample from Cree company reached $\sim$40 $\mu$s \cite{koehl2011room}, indicating that $T_2$ can be improved upon optimizing the sample. The longitudinal spin relaxation time $T_1$ was also measured. The result is shown in Fig. \ref{coherent control at room temperature}(f) for $230 \pm 53$ $\mu$s, which is longer than that of the divacancies in SiC at room temperature \citep{koehl2011room}. This indicates that we can extend $T_2$ to be longer time ($\sim T_1$) using the CPMG-$n$ secquence \cite{naydenov2011dynamical}.

\begin{figure}[!h]
  \centering
  \includegraphics[width=0.6\textwidth]{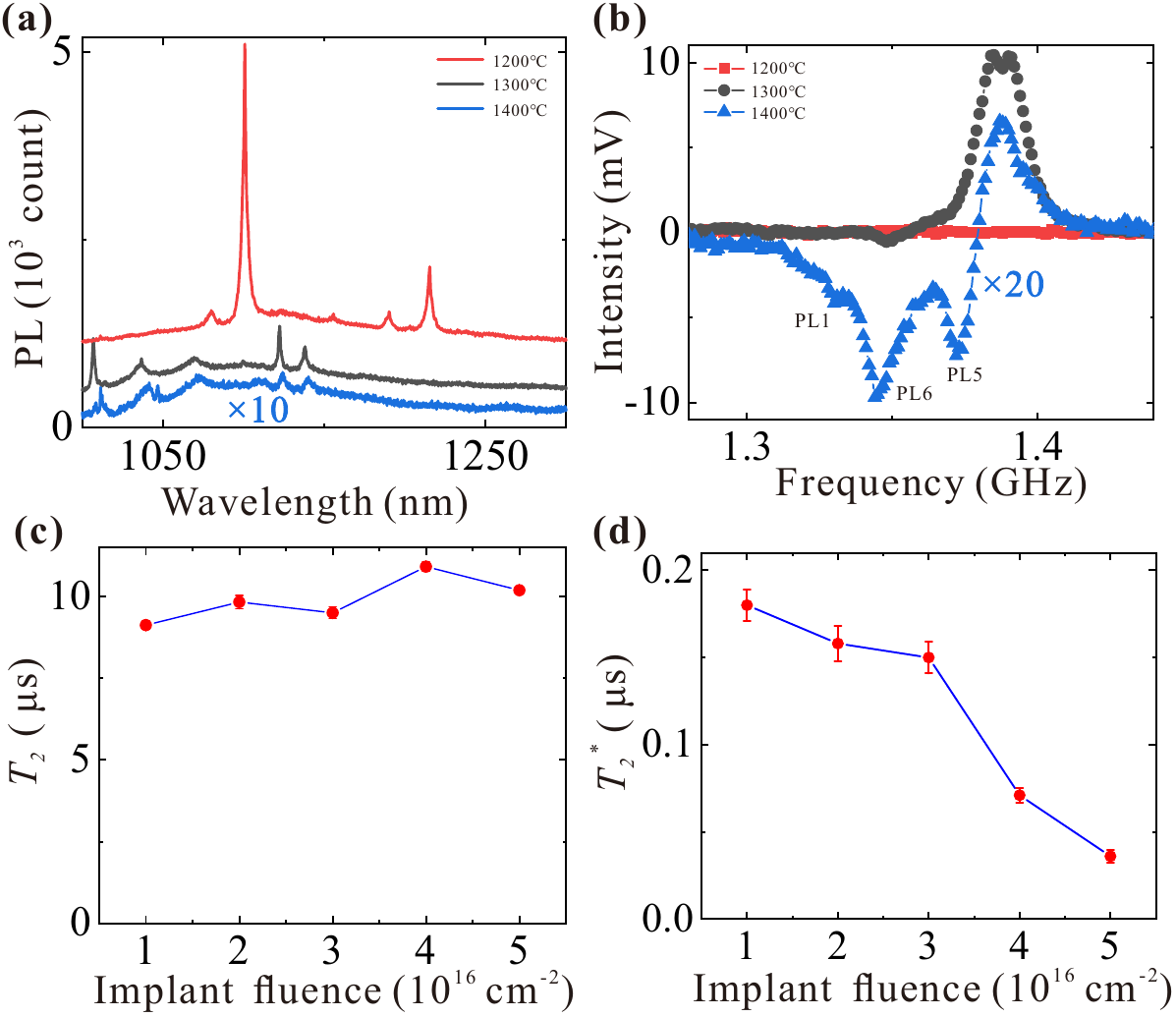}\\

\caption{\textbf{Optical and spin properties of samples annealed at different temperatures and implanted with different fluences.} (a) Photoluminescence spectra of samples at 20 K annealed at 1,200$^\circ$C (red line), 1,300$^\circ$C (black line), and 1,400$^\circ$C (blue line) for 2 h. The same implanted fluence (H$^+$, $2 \times 10^{16}$ ions per $\text{cm}^2$) was applied. (b) The corresponding ODMR signals with red squares (1,200$^\circ$C annealing), black dots (1,300$^\circ$C annealing), and blue triangles (1,400$^\circ$C annealing) at room temperature. There were no detectable ODMR signals of PL8 defects for 1,200$^\circ$C annealing. (c) Room temperature $T_2$ (at 157 G) of PL8 defect spins as a function of the implanted fluence and annealing at 1,300$^\circ$C. (d) Room temperature $T_2^*$ (at 157 G) as a function of the implanted fluence and annealing at 1,300$^\circ$C, which decreased as the implanted fluence increased. Error bars represent the standard deviations of the corresponding fitting values for (c) and (d).}

\label{prepared at different condition}

\end{figure}

Further, we investigated the conditions required to generate PL8 defects. Figs. \ref{prepared at different condition}(a) and \ref{prepared at different condition}(b) show the optical spectra at 20 K and ODMR spectra at room temperature for the samples annealed at 1,200$^\circ$C (red line and squares), 1,300$^\circ$C (black line and dots), and 1,400$^\circ$C (blue line and triangles). All annealing was implemented in a high-vacuum environment ($1.1 \times 10^{-6}$ Pa) for 2 h. The samples were implanted with the same fluence of $2 \times 10^{16}$ cm$^{-2}$ using 170-keV $\text{H}^+$. We found that PL8 defects were very sensitive to the annealing temperature. For the sample annealed at 1,200$^\circ$C, ZPL and ODMR signals of PL8 defects were not observed; however, they were clearly detected when the sample was annealed at 1,300$^\circ$C. When the annealing temperature increased to 1,400$^\circ$C, both optical and ODMR signals considerably reduced. Signals of divacancy defects were also observed in the sample annealed at 1,300$^\circ$C, but they were considerably weaker than those of PL8 defects. Therefore, the controllable generation of PL8 defects in 4H-SiC can be realized by optimizing the annealing temperature. The spin coherent properties of PL8 defects were further investigated as a function of the implanted fluences. The results of $T_2$ and $T_2^*$ are shown in Figs. \ref{prepared at different condition}(c) and \ref{prepared at different condition}(d), respectively. $T_2$ was nearly constant, while $T_2^*$ decreased as the implanted fluences increased owing to inhomogeneous broadening. In SI \cite{SI}, we further measured the optical decay time, which was deduced to be $6.0 \pm 0.1$ ns. By including the spin initialization and measuring the subsequent lifetime, we determined the spin polarization to be approximately 97.5$\%$ (see Supplementary Fig. 1 in SI \cite{SI}).

\section{Discussion}

In this study, we demonstrated a type of defect spin in 4H-SiC which has not been previously reported by implanting high-fluence H$^+$ and annealing at 1,300$^\circ$C. The photoluminescence and ODMR spectra of PL8 defects were different from those of any known defects in 4H-SiC (the optical spectrum excited by 532-nm laser is shown in Supplementary Fig. 2 in SI \cite{SI}) \cite{koehl2011room,ruhl2018controlled}. To identify the possible origins of PL8 defects, we implanted the 4H-SiC sample with high-fluence helium ions (He$^+$) and detected the same ODMR signals as those of PL8 defects (see Supplementary Fig. 3 in SI \cite{SI}). Thus, we conclude that they were not hydrogen-related defects. Other possible candidates were multivacancy \cite{iwata2016density}, stacking faults \citep{ivady2019enhanced}, and silicon vacancy-related defects (Si$_\text{V}^0$ and Si$_\text{V}^{2-}$) with the same spin state as the neutral silicon vacancy in diamond \citep{rose2018observation,weber2010quantum}. Moreover, further theoretical and experimental researches are required to confirm PL8 defect structures. In this study, the spin coherence time $T_2$ of PL8 defects was equal to that of implanted PL6 defects in our sample (see Supplementary Fig. 3 in SI \cite{SI}). That is, we could extend $T_2$ by optimizing the sample \cite{koehl2011room}. As the implanted fluence increased, the concentration of PL8 defects with thermal stability for harsh conditions can be increased to a level that is sufficiently high to be used for maser development \cite{breeze2018continuous}.
\section{Methods}

\noindent
{\textbf {Sample preparation.} The sample used herein was a commercially available high-purity 4H-SiC epitaxy layer purchased from Xiamen Powerway Advanced Material Co. Ltd. The epitaxy thickness was approximately 7 $\mu$m, and the N-doping was $5 \times 10^{15}$ cm$^{-3}$. The epitaxy wafer was implanted with 170-keV H$^+$, and the fluence ranged from $1\times10^{16}$ cm$^{-2}$ to $5\times10^{16}$ cm$^{-2}$. The sample was annealed at varying high temperatures (1,200$^\circ$C, 1,300$^\circ$C, and 1,400$^\circ$C) for 2 h. For comparison, the sample was also implanted with high-fluence He$^+$ ($5\times10^{15}$ cm$^{-2}$ and $2\times10^{16}$ cm$^{-2}$) and carbon ions (C$^{+}$) ($1\times10^{14}$ cm$^{-2}$). The experimental results are shown in SI \cite{SI}.}
\\

\noindent
{\textbf {Experimental setup.} A 920-nm laser modulated by an acousto-optic modulator was used to excite these defects. The laser was reflected by a dichroic mirror (Semrock, Di02-R980) and passed through an objective (N.A. = 0.85 for room-temperature measurement and 0.65 for low-temperature measurement) and focused on the epitaxy SiC sample. The photoluminescence was collected using the same objective and detected by a femtowatt photoreceiver (OE-200-IN1) or spectrometer (Horiba, iHR550) after a 1,000-nm longpass interference filter (Thorlabs, FELH1000). In the ODMR experiment, the microwave was applied using a 20-$\mu$m-diameter copper wire post modulation by 70-Hz pulses from a lock-in amplifier (SRS, SR830). The different PL intensity ($\Delta$PL) between the cases with and without resonant microwave excitation was extracted. The spin signals, which were modulated by 70 Hz, were extracted by the same lock-in amplifier. A microwave generator (mini-circuit, SSG6000) was used to generate the microwave, which was further amplified using a microwave amplifier (mini-circuit, ZHL-30W-252+). The pulse sequences used in the experiment were controlled through a pulse generator (PBESR-PRO-500, Spincore) via a computer.}

\section{Data availability}

The data that support the plots within this paper and other findings of this study are available from the corresponding author upon reasonable request.

\section{Acknowledgements} 
We thank Prof. J\"org Wrachtrup for the helpful discussion and Prof. Jiangfeng Du for assistance of sample annealing. This work was supported by the National Key Research and Development Program of China (Grant No. 2016YFA0302700, 2017YFA0304100, 2017YFE0131300 and 2017YFA0305000), the National Natural Science Foundation of China (Grants No. U19A2075, 61725504, 61905233, 61851406, 61874128, U1732268, 11975221, 11804330, 11821404 and 11774335), the Key Research Program of Frontier Sciences, Chinese Academy of Sciences (CAS) (Grant No. QYZDY-SSW-SLH003, QYZDY-SSW-JSC032), Science Foundation of the CAS (No. ZDRW-XH-2019-1), K.C.Wong Education Foundation(GJTD-2019-11), Anhui Initiative in Quantum Information Technologies (AHY060300 and AHY020100), the Fundamental Research Funds for the Central Universities (Grant No. WK2030380017 and WK2470000026). A.G. acknowledges the National Excellence Program of Quantum-Coherent Materials Project (Hungarian NKFIH Grant No. KKP129866), the EU QuantERA Nanospin Project (Grant No. 127902), the EU H2020 FETOPEN project QuanTelCO (Grant No. 862721), the National Quantum Technology Program (Grant No. 2017-1.2.1-NKP-2017-00001) and the NVKP project (Grant No. NVKP-16-1-2016-0043).

\section{Competing interests}

The authors declare that they have no competing financial interests.

\section{Author contributions}

F.-F. Y. and  A.-L. Y. contributed equally to this work. F.-F. Y. and J.-S. X. designed experiment. A.-L. Y., J.-X. Z. and X. O. prepared the sample. F.-F. Y., P. Y. ,Y. W. and J. D. perform the annealing process. F.-F. Y. carried out the experiment assisted by J.-F. W. and Q. L.. A. G. provided the theoretical support. J.-S. X., X. O., C.-F. L. and G.-C. G. supervised the project. F.-F. Y., A.-L. Y and J.-S. X. wrote the paper with input from other authors. All authors discussed the experimental procedures and results.

\section{Additional information}

Correspondence and requests for materials should be addressed to J.-S.X., X.O. or C.-F.L.

\section{Figure Legends}

\end{document}